\documentclass[showpacs, twocolumn]{revtex4-1}
\usepackage{graphicx}
\usepackage{amsmath}
\usepackage{appendix}

\begin{document}

\title{Quantum dilute droplets of dipolar bosons at finite temperature}

\author{Abdel\^{a}ali Boudjem\^{a}a}
\affiliation{Department of Physics,  Faculty of Exact Sciences and Informatics, Hassiba Benbouali University of Chlef P.O. Box 151, 02000, Ouled Fares, Chlef, Algeria.}
\email {a.boudjemaa@univ-chlef.dz}


\begin{abstract}
We systematically study the properties of dipolar Bose gases with two- and three-body contact interactions at finite temperature
using the Hartree-Fock-Bogoliubov-Popov  approximation.
In uniform case, we obtain an exciting new extension of the seminal Lee-Huang-Yang  corrected equation of state 
that depends explicitly on the thermal fluctuations and on the coupling constant of the three-body interaction. 
We investigate, on the other hand, the effects of thermal fluctuations on the occurrence and stability
of a droplet state in a Bose-Einstein condensate  with strong dipole-dipole interactions.  
We find that at finite temperature, the droplet phase appears as a narrow peak surrounded by a broader thermal halo.
We show that the number of particles inside the droplet decays with increasing temperature.

\end{abstract}


\maketitle

\section{Introduction}

Successful realization and studies of Bose-Einstein condensate (BEC) with dipole-dipole interactions (DDI) which establish a long-range and anisotropic interaction
among particles, inaccessible to a short-range BEC, bring new possibilities to explore novel quantum phase transitions (see for review \cite{Baranov, Pfau,Carr, Pupillo2012}). 
Ultracold dipolar gases have specially attracted much attention, including experiments on magnetic atoms \cite{Gries, Lu, Aik, Paz}, polar molecules \cite{KK, Yan, Tak, Park}, 
Rydbergdressed atoms \cite{Balw}. Ground-state and excited-state  properties of such systems have also been extensively explored 
(see e.g. \cite {Santos, Santos1, Dell, Eberlein, Bon, Corm, Biss, lime, Boudj, Boudj1, Boudj2, Pedri, Petrov1}).

Recently, exquisite experiments of ${}^{164}$Dy atoms made in Stuttgart group  \cite{Pfau1, Pfau2} have shown that when the condensate 
is quenched into a strong DDI regime,  the system instead of collapsing \cite{Lahy}, 
gets into a stable droplet crystal due to the quantum Rosensweig instability \cite {Cow}.
This droplet state is actually characterized by : (i) a large peak density that is only destroyed in a long time scale by three-body losses 
(ii) a decrease in the compressibility of the system \cite{Pfau2}. 
From the theoretical side, two scenarios were performed to explain the stability of such a droplet phase. 
The first scenario based on the presence of a large repulsive three-body interaction \cite{Bess1, Kui, Blakie}.
In the second mechanism, the instability can be halted by quantum fluctuations \cite{Pfau2, Wach, Saito, Bess2, Wach1}.
A similar mechanism has been recently proposed to stabilize droplets in attractive Bose-Bose mixtures \cite{Petrov2}.
These dipolar droplets remain stable even in the absence of external harmonic confinement, forming self-bound ensembles \cite{Wach1, Bess3, Pfau3}.
Most recently, the observation of a macro-droplet state in an ultracold bosonic gas of erbium atoms with strong dipolar interactions has been reported by
the Innsbruck team \cite {Chom}.

In the dilute regime, the dynamics of a stable droplet at zero temperature is generally described with the non-local Gross-Pitaevskii (GP) equation
in which the collapse induced by the attractive mean-field term $\propto n({\bf r})$ ($n({\bf r})$ is the gas density), is arrested by 
the effective repulsive beyond mean-field Lee-Huang-Yang (LHY) term $\propto n^{3/2}({\bf r})$ \cite{Pfau2, Wach, Saito, Bess2, Wach1, Bess3,Chom}.
This term which accounts for the first-order correction to the condensate equation of state (EoS), was originally predicted for a contact-interacting gas \cite {LHY}.


However, up to now, there is a little or no evidence for the finite-temperature effects on this novel state of matter.
It is convenient to remind that experiments actually take place at finite temperatures where the condensate coexists with the thermal cloud.
Effects of this latter become non-negligible as the temperature approaches to the transition and hence, may influence the dynamics and the thermodynamics of the dipolar droplet.
Furthermore, interactions between condensed and noncondensed particles may induce strong thermal fluctuations causing to depopulate the droplet.
These thermal fluctuations could play also a crucial role in the droplet lifetime.

Our goal in this paper is to study, for the first time to our knowledge, the temperature dependence of the droplet state in a dipolar BEC
by profiting of the wealth of the Hartree-Fock-Bogoliubov-Popov (HFBP) theory relies on numerical simulations.
This theory which is gapless,  was used in several early studies to calculate the collective modes and to analyze the thermodynamic properties of both short range and dipolar Bose gases
(see e.g. \cite {Franc, Hut, Bon, Corm, Biss, Tick}).
In the weakly interacting regime, the HFBP as all mean-field theories cannot explain the observed quasi-crystalline droplet patterns  \cite{Pfau1, Pfau2} 
owing to the well known mean-field  collapse.
We show that at sufficiently low temperature, robust droplets require including in addition to the standard LHY correaction, 
a new extra term $\propto n_c^{-1/2}  ({\bf r}) \,T^2$, coming from the thermal fluctuations to the extended GP equation that arrests the dipolar collapse at high condensed density $n_c$.
We reveal that this additional term leads also to shift the validity criterion of the theory.
The HFBP theory within such a generalized LHY (GLHY) corrected EoS enable us to revolutionize our understanding of droplets at nonzero temperatures  since
the thermal fluctuations which emerge naturally are treated on the same footing as the quantum fluctuations.

The rest of the paper is organized as follows. In Sec \ref{3BT}, we will introduce the finite-temperature HFBP model 
for trapped dipolar Bose gases with two- and three-body interactions. 
In Sec.\ref{Unif}, we look at excitations of homogeneous gas and derive useful analytical expressions
for the quantum and thermal fluctuations that depend on the two-body contact interaction, the DDI 
and the coupling constant of the three-body interaction. 
We demonstrate that the peculiar interplay of these quantities  provides a GLHY EoS, and enhances the sound velocity, ground state energy, compressibility 
and the superfluid fraction.
The validity criterion of the theory will be also established.
In Sec.\ref{Drop}, we extend the GLHY result to a spatially inhomogeneous dipolar Bose gas using the local density approximation (LDA).
We then deal with the effects of thermal fluctuations on the nucleation and stability of droplets
at finite temperature by numerically solving the Popov equation in the presence of GLHY stabilization.
The temperature dependence of particles number of the droplet will be also highlighted.
Our conclusions are drawn in Sec.\ref{Conc}.


\section{Three-body model for dipolar bosons} \label{3BT}

We consider a three-dimensional (3D) dilute dipolar Bose gas with contact repulsive  two- and three-body interactions confined in
an external potential $U(r)$. Assuming that the dipoles are oriented perpendicularly to the plane.
The Hamiltonian of the system reads:
\begin{align}\label{ham}
&\hat H = \int d {\bf r} \, \hat \psi^\dagger ({\bf r}) \left(\frac{-\hbar^2 }{2m}\Delta+U({\bf r})\right)\hat\psi(\bf{r}) \nonumber \\
&+\frac{g_2}{2}\int d {\bf r} \hat\psi^\dagger({\bf r}) \hat\psi^\dagger ({\bf r}) \hat\psi({\bf r}) \hat\psi(\bf{r}) \nonumber \\
&+\frac{1}{2}\int d {\bf r} \int d{\bf r'}\, \hat\psi^\dagger({\bf r}) \hat\psi^\dagger ({\bf r'}) V_d({\bf r-r'})\hat\psi({\bf r'}) \hat\psi(\bf{r}) \nonumber \\
&+\frac{g_3}{6}\int d {\bf r} \, \hat\psi^\dagger({\bf r}) \hat\psi^\dagger ({\bf r}) \hat\psi^\dagger ({\bf r}) \hat\psi(\bf{r})  \hat\psi(\bf{r}) \hat\psi(\bf{r}),
\end{align}
where $\hat\psi^\dagger$ and $\hat\psi$ denote, respectively the usual creation and annihilation field operators, $m$ is the particle mass,
$g_2$ and $g_3$ account for the two- and three-body  coupling constants, respectively.
The DDI potential is $V_d({\bf r}) = C_{dd} (1-3\cos^2\theta) / (4\pi r^3)$,
where the coupling constant $C_{dd} $ is ${\cal M}_0 {\cal M}^2$ for particles having a permanent magnetic dipole moment ${\cal M}$ (${\cal M}_0$ is the magnetic permeability
in vacuum) and $d^2/\epsilon_0$ for particles having a permanent electric dipole $d$ ($\epsilon_0 $ is the permittivity of vacuum),
and $\theta$ is the angle between the relative position of the particles ${\bf r}$ and the direction of the dipole.
The two-body coupling constant is defined by $g_2=4\pi \hbar^2 a/m$ with $a$ being  the $s$-wave scattering length which can be adjusted 
using a magnetic Feshbach resonance \cite{Frish,Maier}.
The three-body coupling constant $g_3$ is in general a complex number with $Im(g_3)$ describing the three-body recombination loss
and $Re(g_3)$ quantifying the three-body scattering parameter.

In studying the system dynamics, we will actually be concerned with the equation of motion of the Bose field operator $\hat\psi(\bf{r})$. 
For the Hamiltonian of Eq.(\ref{ham}), this evolves according to the Heisenberg equation of motion
\begin{align}\label{EM}
i\hbar \frac{\partial \hat\psi({\bf r},t)} {\partial t} &= \left[\hat\psi({\bf r},t), \hat H \right] \nonumber \\
&=\left(\frac{-\hbar^2 }{2m}\Delta+U({\bf r})\right)\hat\psi ({\bf r},t) \nonumber \\
&+g_2 \hat\psi^\dagger ({\bf r},t) \hat\psi ({\bf r},t) \hat\psi ({\bf r},t) \nonumber \\
&+\int d{\bf r'}\, \hat\psi^\dagger ({\bf r'},t) V_d({\bf r-r'})\hat\psi({\bf r'},t) \hat\psi({\bf r}, t) \nonumber \\
&+\frac{g_3}{2} \hat\psi^\dagger ({\bf r},t) \hat\psi^\dagger ({\bf r},t) \hat\psi ({\bf r},t)  \hat\psi ({\bf r},t) \hat\psi ({\bf r},t).
\end{align}
It is convenient now to use the Bogoliubov shift \cite {Bog} for the field operator
\begin{equation} \label{decomp}
\hat\psi({\bf r},t)= \Phi({\bf r},t)+\hat {\bar\psi}({\bf r},t),
\end{equation}
where $\Phi({\bf r},t)$ is the condensate wave function and  $\hat {\bar\psi} ({\bf r},t)$ is the field operator of noncondensed atoms.\\
Substituting Eq.(\ref{decomp}) into Eq.(\ref{EM}) and taking the expectation value of the field operator $\hat \psi ({\bf r},t) $ in such a way that 
$\langle \hat {\bar\psi}({\bf r},t) \rangle=0$, we obtain the following  exact evolution equation for the condensate wavefunction

\begin{widetext}
\begin{align}\label{EM1}
i\hbar \frac{\partial \Phi ({\bf r},t)} {\partial t} &=\left \{\frac{-\hbar^2 }{2m}\Delta+U({\bf r}) + g_2 \bigg[n_c({\bf r},t) +2 \tilde n({\bf r},t) \bigg] 
+\frac{g_3}{2} \bigg [n_c^2({\bf r},t)+ 6n_c ({\bf r},t)\tilde n({\bf r},t) + \tilde m^*({\bf r},t) \Phi^2({\bf r},t) \right. \\
&\left.
+ 3 \langle \hat {\bar\psi} ^\dagger ({\bf r},t) \hat {\bar\psi} ^\dagger ({\bf r},t) \hat {\bar\psi} ({\bf r},t) \rangle \Phi ({\bf r},t)
+6 \langle \hat {\bar\psi} ^\dagger ({\bf r},t) \hat {\bar\psi}  ({\bf r},t) \hat {\bar\psi} ({\bf r},t) \rangle \Phi^*({\bf r},t)  
+3 \langle \hat {\bar\psi} ^\dagger ({\bf r},t) \hat {\bar\psi} ^\dagger ({\bf r},t) \hat {\bar\psi} ({\bf r},t) \hat {\bar\psi} ({\bf r},t) \rangle  \bigg ]  \right \} \Phi ({\bf r},t) \nonumber \\
&+\left [ g_2 \tilde m ({\bf r},t) +\frac{3 g_3}{2} \tilde m({\bf r},t)  n_c ({\bf r},t)+ \frac{g_3}{2} \langle \hat {\bar\psi} ({\bf r},t) \hat {\bar\psi}  ({\bf r},t) \hat {\bar\psi} ({\bf r},t) 
\rangle \Phi^*({\bf r},t) +
g_3\langle \hat {\bar\psi} ^\dagger ({\bf r},t) \hat {\bar\psi} ({\bf r},t) \hat {\bar\psi} ({\bf r},t) \hat {\bar\psi} ({\bf r},t) \rangle \right] \Phi^* ({\bf r},t)\nonumber \\
&+ g_2 \langle \hat {\bar\psi} ^\dagger ({\bf r},t) \hat {\bar\psi}  ({\bf r},t) \hat {\bar\psi} ({\bf r},t) \rangle 
+\frac{g_3}{2} \langle \hat {\bar\psi} ^\dagger ({\bf r},t) \hat {\bar\psi} ^\dagger ({\bf r},t) \hat {\bar\psi} ({\bf r},t) \hat {\bar\psi} ({\bf r},t) \hat {\bar\psi} ({\bf r},t) \rangle \nonumber \\
&+\int d{\bf r'} V_d({\bf r}-{\bf r'}) \bigg\{ \bigg [ n_c ({\bf r'},t) +\tilde n ({\bf r'},t) \bigg] \Phi({\bf r},t)+ \tilde n ({\bf r},{\bf r'},t)\Phi({\bf r'},t) +\tilde m ({\bf r},{\bf r'},t)\phi^*({\bf r'},t) 
\nonumber \\
&+\langle \hat {\bar\psi}^\dagger({\bf r'},t) \hat {\bar\psi}({\bf r'},t) \hat {\bar\psi} ({\bf r},t)\rangle \bigg \}, \nonumber 
\end{align}
\end{widetext}
where $\Phi ({\bf r})=\langle \hat\psi ({\bf r})\rangle$  is the condensate wavefunction, $n_c({\bf r})=|\Phi({\bf r})|^2$ is the condensed density,
$\tilde n ({\bf r})= \langle \hat {\bar\psi}^\dagger ({\bf r}) \hat {\bar\psi} ({\bf r}) \rangle $  is the noncondensed density
and $\tilde m ({\bf r})= \langle \hat {\bar\psi} ({\bf r}) \hat {\bar\psi} ({\bf r}) \rangle $ is the anomalous density which can be interpreted as the density of pair-correlated atoms.
The terms $\tilde n ({\bf r, r'})$ and $\tilde m ({\bf r, r'})$ are, respectively the normal and the anomalous one-body density matrices.
They represent the dipole exchange interaction between the condensed and noncondensed atoms.
The quantities $\langle \hat {\bar\psi} ^\dagger ({\bf r}) \hat {\bar\psi}  ({\bf r}) \hat {\bar\psi} ({\bf r}) \rangle$, 
$\langle \hat {\bar\psi} ^\dagger ({\bf r}) \hat {\bar\psi} ^\dagger ({\bf r}) \hat {\bar\psi} ({\bf r}) \hat {\bar\psi} ({\bf r}) \rangle$
and $\langle \hat {\bar\psi} ^\dagger ({\bf r}) \hat {\bar\psi} ^\dagger ({\bf r}) \hat {\bar\psi} ({\bf r}) \hat {\bar\psi} ({\bf r}) \hat {\bar\psi} ({\bf r}) \rangle $
account, respectively for the third, fourth, and fifth order anomalous correlation functions.\\
Equation (\ref{EM1}) clearly shows that the condensate is dynamically coupled with the noncondensate and anomalous averages.
For $g_3=0$, Eq.(\ref{EM1}) recovers the generalized nonlocal finite temperature GP equation obtained recently in our work \cite {Boudj1} 
using the representative ensembles theory.
For $\tilde n =\tilde m =0$ and for vanishing higher order anomalous correlators, Eq.(\ref{EM1}) reduces to the usual nonlocal GP equation 
which describes dipolar Bose gases only at zero temperature.

The HFBP approximation consists of omitting all terms associated with anomalous correlations and  keeping only $n_c$ and $\tilde n$. This yields
\begin{widetext}
\begin{align}\label{EMP}
i\hbar \frac{\partial \Phi ({\bf r},t)} {\partial t} &=\left \{ h^{sp}+ g_2 \bigg[n_c({\bf r},t) +2 \tilde n({\bf r},t) \bigg] 
+ \frac{g_3}{2}\bigg [n_c^2({\bf r},t)+ 6n_c ({\bf r},t)\tilde n({\bf r},t)  \bigg ]  \right \} \Phi ({\bf r},t) \\
&+\int d{\bf r'} V_d({\bf r}-{\bf r'}) \bigg \{ \bigg [ n_c ({\bf r'},t) +\tilde n ({\bf r'},t) \bigg] \Phi({\bf r},t)+ \tilde n ({\bf r},{\bf r'},t)\Phi({\bf r'},t) \bigg \}, \nonumber 
\end{align}
\end{widetext}
where $h^{sp} =(-\hbar^2/2m) \Delta +U({\bf r})$ is the single particle Hamiltonian.
The main feature of Eq.(\ref{EMP}) is that it ensures important conservation laws, such as number of particles and energy conservation.\\ 
The collective modes or elementary excitations of the system corresponding to Eq.(\ref{EMP}) can be usually found by 
looking for solutions of the form: 
$$\Phi({\bf r},t)=[\Phi_0({\bf r})+\delta \Phi({\bf r},t) ] e^{-i\mu t/\hbar},$$
where $\delta \Phi = \sum_k [u_k ({\bf r}) e^{-i \varepsilon_k t/\hbar}+ v_k({\bf r}) e^{i \varepsilon_k t/\hbar}] $ 
represents fluctuations of the condensate wavefunction around the equilibrium solution $\Phi_0$.
After some algebra, we obtain the generalized non-local Bogoliubov-de-Gennes (BdG) equations
\begin{widetext}
\begin{align} 
\varepsilon_k u_k ({\bf r}) &= \hat {\cal L} u_k ({\bf r})+ \hat {\cal M} v_k ({\bf r}) + \int d{\bf r'} V_d({\bf r}-{\bf r'}) n ({\bf r},{\bf r'}) u_k ({\bf r'}) 
+ \int d {\bf r'} \Phi_0({\bf r'})  V_d({\bf r}-{\bf r'}) \Phi_0({\bf r})  v_k ({\bf r'}), \label{BdG1} \\ 
-\varepsilon_k v_k ({\bf r}) &= \hat {\cal L} v_k ({\bf r})+ \hat {\cal M} u_k ({\bf r}) + \int d{\bf r'} V_d({\bf r}-{\bf r'}) n ({\bf r},{\bf r'}) v_k ({\bf r'}) 
+ \int d {\bf r'} \Phi_0({\bf r'})  V_d({\bf r}-{\bf r'}) \Phi_0({\bf r})  u_k ({\bf r'}), \label{BdG2}
\end{align}
\end{widetext}
where 
$\hat {\cal L}=h^{sp}+ 2g_2n ({\bf r})+ 3 g_3 [n_c^2 ({\bf r}) +4 n_c \tilde n ({\bf r})]/2 + \int d{\bf r'} V_d({\bf r}-{\bf r'}) n ({\bf r'})-\mu$
in which $n({\bf r})=n_c({\bf r})+\tilde n ({\bf r})$ is the total density, $\hat {\cal M}=g_2 \Phi_0^2({\bf r})+ g_3[ n_c^2({\bf r})+3\Phi_0^2({\bf r}) \tilde n ({\bf r}) ]$,
$n ({\bf r},{\bf r'})= \Phi_0^*({\bf r'}) \Phi_0({\bf r})+ \tilde n ({\bf r},{\bf r'})$.\\
The solution of Eqs.(\ref{BdG1}) and (\ref{BdG2}) gives the Bogoliubov eigenfrequencies  $\varepsilon_k$ and the corresponding
eigenfunctions $ u_k({\bf r}), v_k({\bf r}) $ of the excitations, which obey the normalization condition
$$\int d {\bf r} [u_k^*({\bf r}) u_{k'}({\bf r})- v_k^*({\bf r}) v_{k'}({\bf r})]=\delta_{kk'}.$$
Note that the BdG equations (\ref{BdG1}) and (\ref{BdG2}) can also be derived by an alternative procedure of diagonalizing the above Hamiltonian, 
in which one expresses the noncondensed field operator $\hat {\bar\psi}=\sum_k [u_k ({\bf r}) \hat b_k+ v_k^*({\bf r}) \hat b_k^\dagger] $ 
in terms of  the bosonic quasiparticles annihilation and creation operators $\hat b_k$ and $\hat b_k^\dagger$, as discussed in \cite {Bon,Corm}.

The equilibrium thermal one-body density matrix is given by
\begin{align} \label{Nconds}
 \tilde n ({\bf r, r'})&= \sum_k \bigg\{ \left[u_k^*({\bf r'}) u_k ({\bf r})+v_k({\bf r'})v_k^*({\bf r}) \right] N_k({\bf r}) \\
&+v_k({\bf r'})v_k^*({\bf r})\bigg\},\nonumber 
\end{align}
where $N_k=\langle \hat b_k^{\dagger} \hat b_k\rangle=[\exp(\varepsilon_k/T)-1]^{-1}$ are occupation numbers for the excitations.  The noncondensed density can simply  be found 
by putting $  \tilde n ({\bf r})=\tilde n ({\bf r, r})$ in Eq.(\ref{Nconds}).

For computational feasibility, we assume that $\tilde n ({\bf r},{\bf r'})=0$ for ${\bf r} \neq {\bf r'}$ in Eqs.(\ref{EMP})-(\ref{Nconds}) \cite{Bon}.
One should stress that the neglect of the long range exchange term $\int d{\bf r'} V_d({\bf r}-{\bf r'}) n ({\bf r},{\bf r'}) u_k ({\bf r'}) $ does not qualitatively affect
the stability of the system \cite {Bon, Biss}.
The study of the thermodynamics of a trapped dipolar Fermi gas showed that the exchange terms are practically less important than the direct ones \cite {Zhan, Bail}. 
Recent finite-temperature analysis of a quasi-2D dipolar gas based on the HFBP has revealed also that these exchange terms are not important \cite {Tick}.
Another justification of the above approximation is that at high temperature where the HFBP approach is valid,
the correlation function (dipole thermal exchange) $\tilde n ({\bf r},{\bf r'})$ goes to zero.


\section{Uniform case}  \label{Unif}

In this section we set up the necessary theory to compute self-consistently the condensate fluctuations, elementary excitations and some thermodynamic quantities
of a dipolar homogeneous ($U({\bf r})=0$) Bose gas in the presence of the three-body interactions. 
The field operator of noncondensed particles transforms as
$\hat {\bar\psi} ({\bf r})= (1/V) \sum_{\bf k} \hat a_k e^{i {\bf k}. \bf r}$ with $V$ being the system volume, 
and the DDI potential takes the form $\tilde V_d ({\bf k})=C_{dd} (3\cos^2\theta_k-1)/3 $ \cite {Baranov, Boudj}, 
where the vector ${\bf k}$ represents the momentum transfer imparted by the collision. 

Assuming the weakly interacting regime where  $ r_* \ll  \xi$ with $r_*=m C_{dd}/4\pi \hbar^2$ 
being the characteristic dipole-dipole distance and  $\xi=\hbar^2/\sqrt{mn_cg_2 \left(1+g_3n_c/g_2\right)}$ is the healing length of BEC with three-body interactions. \\
The chemical potential can be easily obtained from Eq.(\ref{EMP}) as
\begin{equation} \label{chim}
\mu= g_2 [n_c +2 \tilde n] + \frac{g_3}{2} [n_c^2+6n_c\tilde n] +\tilde V_d (0) n,
\end{equation}
Substituting (\ref{chim}) into (\ref{BdG1}) and (\ref{BdG2}), we obtain  the following useful Bogoliubov  three-body dispersion relation
\begin{equation} \label{spec}
\varepsilon_k= \sqrt { \omega_k^2-\Delta_k^2},
\end{equation}
where 
$$\omega_k= \frac{\hbar^2k^2}{2m} +2g_2n+ \frac{3g_3}{2} [n_c^2 +4 n_c \tilde n]+\tilde V_d (0) n +\tilde V_d ({\bf k}) n_c -\mu$$
and 
$$\Delta_k= g_2 n_c+ g_3 [n_c^2 +3 n_c \tilde n]+ \tilde V_d ({\bf k}) n_c.$$
In the low momenta limit ($k\rightarrow 0$), the spectrum (\ref{spec}) is a sound wave $\varepsilon_k= \hbar c (\theta) k$ \cite{Boudj, Boudj1} where
 the sound velocity is given by
\begin{align} \label{soundd}
c (\theta) &= \Delta (0)/m \\ \nonumber
&= c_0 \sqrt{ (1+g_3n_c/g_2) [1+\gamma (3\cos^2\theta-1)]} \,,
\end{align}
where $c_0=\sqrt{g_2n_c /m}$ is the zeroth order sound velocity, and the dimensionless parameter $\gamma$ is  given by
\begin{equation} \label{PG}
\gamma= \epsilon_{dd}/(1+g_3 n_c/g_2), 
\end{equation}
where $\epsilon_{dd}=C_{dd}/3g_2 $ can be computed from the background scattering length.
For instance,  for ${}^{164}$Dy atoms, $\epsilon_{dd} = 1.45$ \cite{Pfau2} and for ${}^{166}$Er atoms, $\epsilon_{dd} = 0.8$ \cite{Chom}.
Equation (\ref{soundd}) clearly shows that the sound velocity is modified by three-body forces which may lead to enhance the collective modes of the system.
We see also that $c (\theta)$ acquires a dependence on the propagation direction, which is fixed by the angle $\theta$. 
The anisotropy of the sound velocity has been already achieved experimentally using the Bragg spectroscopy technique few years ago \cite {bism}.

The noncondensed density can be calculated using the Fourier transform of (\ref{Nconds}), we find

\begin{align}\label {dep}
\frac{\tilde{n}}{n_c}=&\frac{ 8}{3} \sqrt{\frac{ n_c a^3}{\pi}}  \left(1+ \frac{g_3n_c}{g_2}\right)^{3/2} h_3(\gamma) \\
&+\frac{2}{3}\sqrt{\frac{n_c a^3}{\pi}} \left(\frac{\pi T}{\mu_0}\right)^{2} \left(1+ \frac{g_3n_c}{g_2}\right)^{-1/2} h_{-1}(\gamma), \nonumber
\end{align}
where $\mu_0=g_2n_c$ and the functions $$h_j(\gamma)=(1-\gamma)^{j/2} {}_2 F_1\left(-\frac{j}{2},\frac{1}{2};\frac{3}{2};\frac{3\gamma}{\gamma-1}\right)$$
represent the contribution of the DDI into the condensate depletion, ${}_2F_1$ is the hypergeometric function.
The leading term in Eq.(\ref{dep}) represents the quantum fluctuations. 
The last term which accounts for the effect of the thermal fluctuations \cite {Boudj1}, is calculated at temperatures $T\ll g_2n_c $, 
where the main contribution to (\ref{Nconds}) comes from the region of small momenta ($\varepsilon_k=\hbar c (\theta) k$).
At higher temperatures i.e. $T \gg g_2 n_c$, the main contribution to (\ref{Nconds}) comes from the single particle excitations. 
Therefore, the thermal contribution of $\tilde{n}$ becomes identical to the density of noncondensed atoms in an ideal Bose gas \cite {Boudj1,Boudj2}.

For $\gamma>1$ or equivalently $\epsilon_{dd} > (1+g_3 n_c/g_2)$, the functions $h_j(\gamma)$ become imaginary. 
In this case, the dipolar interaction, which is partially attractive, dominates the repulsive two- and three-body contact interactions leading to the collapse of the condensate
results in from the presence of unstable soft modes. 
In the absence of the three-body interactions $g_3=0$, expression (\ref {dep}) reduces to that obtained for a dipolar BEC with two-body interactions \cite{Boudj}.
For a condensate with pure contact interactions ($h_3(\epsilon_{dd}=0)=1$), $\tilde{n}$  takes the from:
\begin{align}\label {dep1}
\frac{\tilde{n}}{n_c}=&\frac{ 8}{3} \sqrt{\frac{ n_c a^3}{\pi}} \left(1+ \frac{g_3n_c}{g_2}\right)^{3/2}  \nonumber\\
&+\frac{2}{3}\sqrt{\frac{n_c a^3}{\pi}} \left(\frac{\pi T}{\mu_0}\right)^{2} \left(1+ \frac{g_3n_c}{g_2}\right)^{-1/2}. \nonumber
\end{align}
Near the collapse threshold i.e. $\epsilon_{dd}\approx 1+g_3 n_c/g_2$,  $ h_3(\gamma) \simeq 1.3$, the condensed depletion 
exceeds its value of a pure contact two-body interaction
which means that both the DDI and the three-body interactions may enhance the quantum and thermal fluctuations of the condensate.
At $T=0$ and for $g_3=0$, Eq.(\ref {dep}) reproduces formally the results of Ref \cite{lime}.

One can infer that the small parameters of the theory require the inequalities

\begin{equation}\label{VC}
\begin{cases}
\sqrt{ n_c a^3} (1+g_3n_c/g_2)^{3/2} h_3(\gamma) \ll 1, & T=0\\[1.5ex]
\frac{T}{\mu_0}\sqrt{ n_c a^3}  (1+g_3n_c/g_2)^{-1/2} h_{-1}(\gamma) \ll 1. & T \ll g_2 n_c\\[1.5ex]
\end{cases}
\end{equation}
The conditions (\ref{VC}) differ by the  factors $(1+g_3n_c/g_2)^{j/2} h_j(\gamma)$ from the standard small 
parameters of the theory in the absence of the DDI and the three-body interaction. 
If the DDI vanishes $( h_3(\epsilon_{dd}=0)=1)$, the validity criterion of the theory becomes
$\sqrt{ n_c a^3} (1+g_3n_c/g_2)^{3/2} \ll 1$ at $T=0$, and $(T/g_2n_c)\sqrt{ n_c a^3}  (1+g_3n_c/g_2)^{-1/2} \ll 1$ at $T \ll g_2 n_c$.

Corrections to the chemical potential due to the quantum and thermal fluctuations are given by \cite{Boudj, Boudj2}
$$\delta \mu=\int \tilde V({\bf k}) [v_k(v_k-u_k)+(v_k-u_k)^2N_k ] d {\bf k}/(2\pi)^3,$$ 
where $ u_k,v_k=(\sqrt{\varepsilon_k/E_k}\pm\sqrt{E_k/\varepsilon_k})/2$. 
This integral is ultraviolet divergent. One way to circumvent such a problem is the use of the dimensional regularization \cite{Boudj1}
which is an accurately defined mathematical procedure in the limit of weak interaction.
From this method follows a useful GLHY corrected EoS \cite{LHY} for a dipolar BEC with three-body interactions
\begin{align}\label {chem}
\frac{\delta\mu} {\mu_0}&=\frac{32}{3} \sqrt{\frac{ n_c a^3}{\pi}} \left(1+ \frac{g_3n_c}{g_2}\right)^{5/2} h_5(\gamma)\\
&+\frac{4}{3}\sqrt{\frac{n_c a^3}{\pi}} \left(\frac{\pi T}{\mu_0}\right)^{2} \left(1+ \frac{g_3n_c}{g_2}\right)^{1/2} h_1(\gamma), \nonumber
\end{align}
Obviously, this equation describes rigorously the effects of quantum and thermal fluctuations on the system at hand.
At $T=0$ and for $g_3=0$, Eq.(\ref{chem}) coincides with that found recently in Refs \cite {Boudj, Boudj1,lime}.
For a condensate with a pure two-body contact interaction ($h_5(\epsilon_{dd}=0)=1$ and $g_3=0$), the above EoS simplifies to the standard  LHY EoS. 
Equation (\ref{chem}) shows that the contribution of the DDI in the thermal part of $\delta \mu$ is less important than that in the quantum part.
This is because the function $h_5(\gamma)$ grows monotonically with $\gamma$ while $h_1(\gamma)$ decreases with $\gamma$ and vanishes for $\gamma \sim 1$. 
Furthermore, Eq.(\ref{chem}) is appealing since it provides extra terms arising from quantum and thermal fluctuations 
which may compensate at large enough densities of the attractive mean field term, leading to a stable droplet. This occurs even in the absence of the 
three-body interactions as we will see in Sec.\ref{Drop}.




At $T = 0$, the inverse compressibility is defined as $\kappa^{-1} = n^2\partial\mu/\partial n$. 
Then, using (\ref{chem}), we get 
\begin{equation}\label {compr}
\frac{ \delta \kappa^{-1}}{g_2}=16\,n_c^2\sqrt{\frac{n_c a^3 }{\pi }}  \left(1+ \frac{g_3n_c}{g_2}\right)^{3/2} {\cal G} (\gamma), 
\end{equation}
where 
\begin{align}
{\cal G} (\gamma)&= \frac{1}{\gamma-1}  \bigg[  \left(\epsilon_{dd} -1-\frac{3g_3n_c}{g_2} \right)  h_5 (\gamma)  
\nonumber\\ 
&
+\frac{g_3n_c} {3g_2} \left(1-\frac{3\gamma}{\gamma-1} \right)^{5/2} \bigg].  \nonumber
\end{align}
For $g_3=0$, this equation reduces immediately to our recent formula obtained for a 3D dipolar BEC with two-body contact interaction \cite{Boudj1}. 
For $\epsilon_{dd} = 0$, $\delta \kappa^{-1}/g_2= 16 n_c^2\sqrt{n_c a^3/\pi}  (1+ g_3n_c/g_2)^{3/2} (1+ 8g_3n_c/3g_2)$.
The inverse isothermal compressibility can be obtained easily from $(\partial P/\partial n)_T$, where $P$ is the pressure of the system. 
We should stress that the compressibility is an important quantity for indicating the transition between the condensate and the droplet phase.

The energy shift due to the quantum and thermal fluctuations can be computed by integrating the chemical potential (\ref{chem}) with respect to the density.

Now let us use our model to analyze the behavior of the superfluid fraction.
In a 3D dipolar BEC the superfluid density $n_s$ is a tensor quantity with components $n_s^{ij}$ due to the peculiar anisotropy property of the DDI \cite{Axel3, Boudj3}.
This means that $n_s$ depends on the direction of the superfluid motion with respect to the orientation of the dipoles.
It can be written as \cite{Boudj3}
\begin{equation} \label{sup1}
 \frac{n_s^{ij}}{n}= \delta_{ij}- \frac{2}{Tn}\int \frac{d {\bf k}}{(2\pi)^3} \left[\frac{\hbar^2}{2m} \frac{k_i k_j}{4\, \text {sinh}^2 (\varepsilon_k/2T)}\right]. 
\end{equation}
At low temperatures $T\ll n_c g_2$, the parallel direction of the superfluid fraction reads
\begin{equation}\label {supflui1}
\frac{n_s^{\parallel}}{n} =1- \frac{2\pi^2 T^4}{45 m n \hbar^3  c_0^5} \left(1+ \frac{g_3n_c}{g_2}\right)^{-5/2} h_{-5}^{\parallel} (\gamma),
\end{equation}
where the functions
$$h_j^{\parallel} (\gamma)=\frac{1}{3} (1-\gamma)^{j/2} {}_2\!F_1\left(-\frac{j}{2},\frac{5}{2};\frac{3}{2};\frac{3\gamma}{\gamma-1}\right),$$ 
behave as: $h_j^{\parallel} (\gamma=0)=1/3$ and imaginary for $\gamma>1$. \\
In the perpendicular direction, one has
\begin{equation}\label {supflui2}
\frac{n_s^{\perp}}{n} =1- \frac{\pi^2 T^4}{45 m n \hbar^3  c_0^5} \left(1+ \frac{g_3n_c}{g_2}\right)^{-5/2} h_{-5}^{\perp}(\gamma),
\end{equation}
where $h_j^{\perp}(\gamma)= h_j(\gamma)-h_j^{\parallel} (\gamma)$. \\
Expressions (\ref{supflui1}) and (\ref{supflui2}) show that the DDI and the three-body interaction may significantly reduce the two different superfluid fractions.
A direct comparison between both components shows that  $n_s^{\parallel}$ and  $n_s^{\perp}$ coincide for $\epsilon_{dd}=0$.
They well reproduce the two-body contact interaction result for $\epsilon_{dd}=0$ and $g_3=0$.
Moreover, we read off from Eqs.(\ref{supflui1}) and (\ref{supflui2}) that for sufficiently large value of $\gamma$, $n_s^{\parallel}$ is smaller than $n_s^{\perp}$
due to the fast decay of  $h_j^{\parallel}$ compared to $ h_j(\gamma)$. 
At high temperature $T\gg n_c g$, the normal part of $n_s$ coincides with the noncondensed density of an ideal Bose gas.
These outcomes could provide important insights into the superfluidity of liquid helium droplets \cite{Dalfovo}.

\section{Quantum droplets}  \label{Drop}

In this section we investigate the role of thermal fluctuations on the formation of quantum droplets
in a strongly dipolar BEC.
Let us consider at this stage a cylindrically symmetric harmonic potential $U({\bf r})=\frac{1}{2}m \omega_{\rho}^2 (\rho^2+\lambda^2 z^2)$,
where $\rho^2=x^2+y^2$, and $\lambda=\omega_{z}/\omega_{\rho}$ is the ratio between the trapping frequencies in the axial and radial directions.
We adopt the LDA  to evaluate the excitation spectrum and quantum fluctuations semi-classically,
setting $ \varepsilon_k \rightarrow \varepsilon_k ({\bf r})$ and $ \delta \mu\rightarrow  \delta \mu({\bf r})$.
Inserting these corrections in the generalized GP equation (\ref{EMP}), we get
\begin{align}\label{LDA}
i\hbar \frac{\partial \Phi ({\bf r})} {\partial t} =\left [ h^{sp}+ \mu ({\bf r})+\delta \mu({\bf r})\right] \Phi ({\bf r}), 
\end{align}
where $\mu ({\bf r})=g_2[n_c({\bf r})+2\tilde{n}({\bf r})]+g_3 [n_c^2({\bf r})+ 6n_c ({\bf r})\tilde n({\bf r})]/2+ \int d{\bf r'} V_d({\bf r}-{\bf r'})  n ({\bf r'})$. \\
The main feature emerging from this equation is that the temperature appears naturally without any subsidiary approximation 
which is not the case in the standard GP equation of Refs \cite{Bess1, Wach}.
The LDA treatment of the GLHY term in Eq.(\ref{LDA}) is applicable when the external potential is sufficiently smooth and $u_k({\bf r})$ and $v_k({\bf r})$ 
are slowly varying functions of the position. 
This is indeed the case of relatively big droplets since the system remains in the Thomas-Fermi regime \cite{Wach}.
Recent Path Integral Monte Carlo calculations  \cite{Saito} and experimental measurement \cite{Chom} have checked that the LDA gives reasonable results.

\begin{figure}[ htb] 
\includegraphics[scale=0.35]{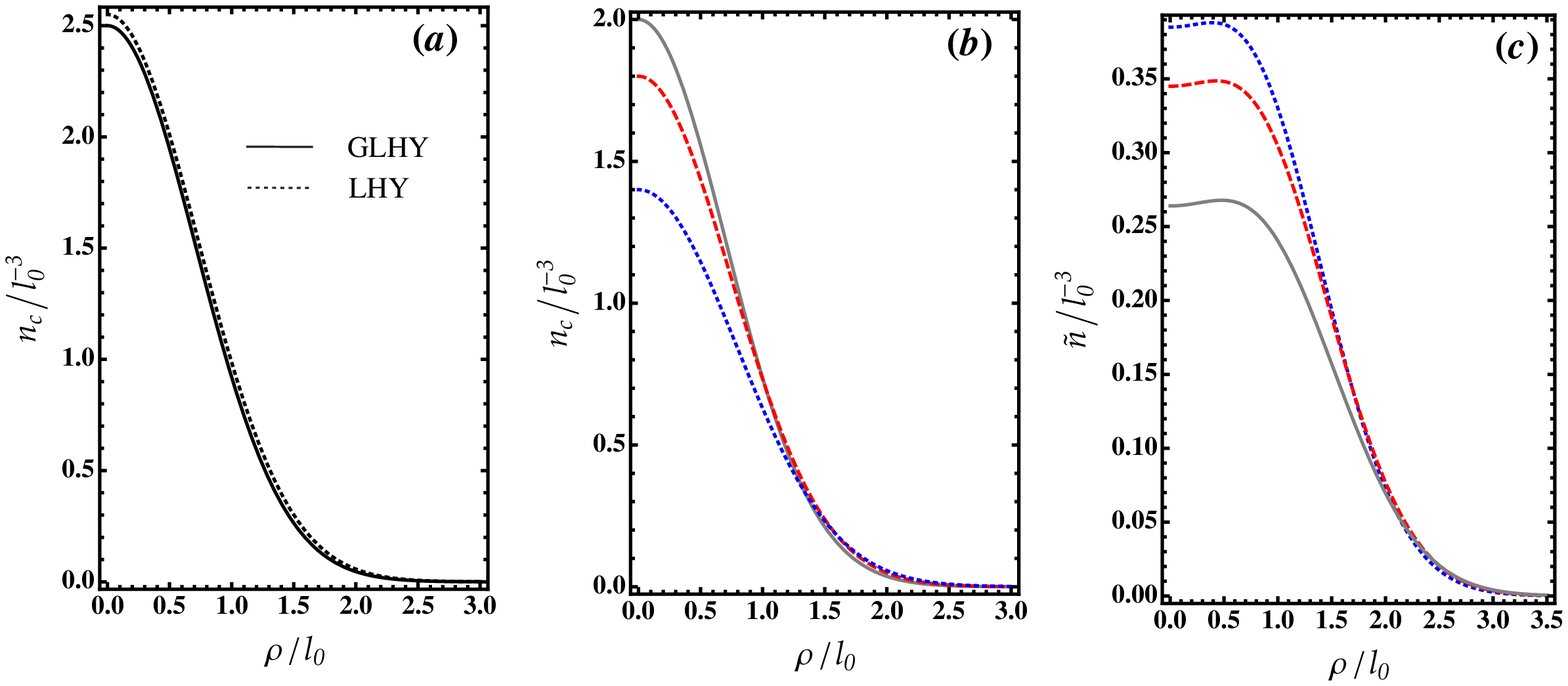}
\caption {(Color online) (a) Density profiles of a droplet state at $T=0$. 
(b) Density profiles of a droplet at several values of temperature. 
(c) Density of the thermal halo at various temperatures.
Parameters are : $N=15\times 10^3$ of ${}^{164}$Dy  atoms, $\omega_{\rho}=2\pi \times 45$ Hz, $\lambda=3$, the  scattering length $a=80 a_0$ and $g_3=0$ \cite{Pfau1}.
Gray solid lines: $T=0.2 \,T_c^0$, dashed red lines: $T=0.4 \,T_c^0$ and dotted blue lines: $T=0.65 \,T_c^0$.
Here $T_c^0= (N/\zeta (3))^{1/3}  \hbar \bar\omega $, is the ideal gas critical temperature, where $\bar\omega = (\omega_{\rho}^2\omega_z)^{1/3}$ is
the geometric mean of the trap frequencies.}
\label{Ddens}
\end{figure}

\begin{figure}[ htb] 
\includegraphics[scale=0.35]{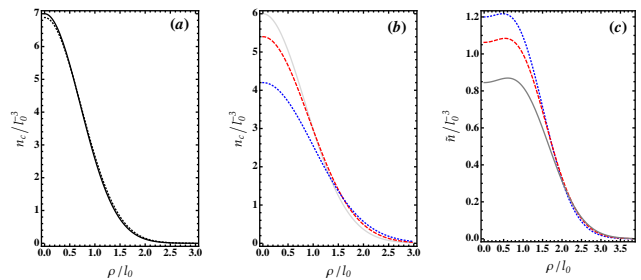}
\caption {(Color online) The same as Fig.\ref{Ddens} but for Er droplets.
Parameters are: $N=1.2\times 10^5$ of ${}^{166}$ Er atoms, $\omega_{\rho}= 2\pi \times 177$ Hz, $\lambda \simeq 0.1$, 
the relative scattering length $\epsilon_{dd}=0.8$ and $Im (g_3)=3.7 \times 10^{-42}$ m$^6/$s \cite{Chom}.
Here we adopt the method described in Refs \cite {Blakie, Bess3} for accounting three-body losses. }
\label{Ddens1}
\end{figure}

To benchmark the GLHY formalism, we solve iteratively and selfconsistently  Eqs.(\ref{LDA}), (\ref{BdG1})-(\ref{Nconds}), 
and carry out a quantitative comparison to the standard LHY approximation used in Refs \cite {Pfau2, Bess1, Kui, Wach, Blakie, Chom}.
In our numerical simulation, we follow the previous experimental and theoretical works and assume that  
the repulsive three-body interactions are not relevant i.e. $Re(g_3)=0$ \cite{Pfau2, Wach, Saito, Bess2, Wach1} .
In this case the thermal fluctuations of the excitations result in the GLHY correction, lead to an additional repulsive term in Eq.(\ref{LDA}) scale as  
$(4\sqrt{\pi^3a^3}/3g_2)\, h_1(\epsilon_{dd})\, T^2 n_c^{-1/2} ({\bf r})$ at temperatures below  $T_c$.
It is worth noticing that the usual calculation of the LHY correction with two-body interactions requires a summation of Bogoliubov modes up
to some cutoff to prevent an ultraviolet divergence due to the limitation of the pseudopotential. The condensate interaction part
also needs an adjustment in order to take into account this cutoff. 
Therefore, the semiclassical theory must be carried out carefully to evaluate precisely  the effects of quantum fluctuations in quench experiments \cite{Wach}.
The low-momentum cutoff that we introduce in the GLHY correction is of the form $k_c(\theta)=k_z\sqrt{\cos^2\theta+ \lambda^2 \sin^2\theta}$ \cite{Wach}
which can be utilized to obtain  the quantum and thermal fluctuations by numerically evaluating  $h_5(\epsilon_{dd}$) and $h_1(\epsilon_{dd})$. 
Such a cutoff procedure has been shown to be somehow consistent with lowest order expansion of the functions $h_j(\epsilon_{dd}$) 
leading to effectively neglect the imaginary parts of $h_j(\epsilon_{dd}$) for $\epsilon_{dd}>1$ which is very small compared to the real part \cite{Bess2}.

Density profiles obtained with this resolution are displayed in Fig.\ref{Ddens}.
We observe in Fig.\ref{Ddens}.a that even at $T=0$, the GLHY calculations show slightly  lower density compared to the standard LHY theory. 
This discrepancy can be attributed to the quantum depletion which becomes significant for strong DDI (as in ${}^{164}$Dy BEC) even at zero temperature (see Eq. (\ref{dep})).
On the other hand, Fig.\ref{Ddens}.b shows that for growing temperatures some part of atoms seemingly leaves the droplet region
forming a thermal halo-like structure.
As shown in  Fig.\ref{Ddens}.c, the density of such a thermal component which is almost Gaussian in shape, is increasing with temperature. 
For instance, at $T=0.4 \,T_c^0$, the amount of atoms gathered in the halo is approximately 20\%. 
It is clearly visible that the size of the thermal halo is larger compared to that of the droplet  and is strongly dependent on temperature.

The same behavior persists in Fig.\ref{Ddens1} where a macro-droplet state is generated in a dipolar quantum fluid of Er atoms
and remarkable decoupling between the thermal component and the dense core of the droplet is observed  \cite{Chom}.
In this case,  the quantum and thermal fluctuations are strong enough to dominate the three-body losses, leaving the system in the high density
regime and hence,  a single macro-droplet will take place \cite{Wach, Bess2, Chom}.

\begin{figure}[ htb] 
\includegraphics[scale=0.8]{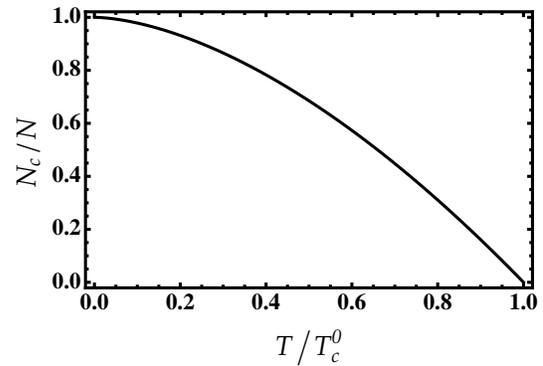}
\caption { Number of particles in the droplet a function of the reduced temperature. 
Parameters are the same as in Fig.\ref{Ddens}.}
\label{TDrop}
\end{figure}

Figure \ref{TDrop} depicts that the number of particles inside the droplet decreases with rising temperature and vanishes close to $T_c$.
This indicates that the droplets (whatever their species; Er, Dy, ...) are individually Bose condensed and superfluid. 
The fact that the droplet is superfluid at $T\ll T_c $ is indeed not surprising, knowing that particles in a droplet are basically very weakly-interacting
due to the depression of the potential at short distance.

\section{Conclusions} \label{Conc}

In this paper we have deeply investigated the properties of a dilute 3D dipolar Bose gas confined by a cylindrically symmetric harmonic trapping potential
in the presence of three-body interactions and losses at finite temperature.

By applying the HFBP formalism, we have derived a generalized equation of motion treating in a self-consistent manner the dynamics of dipolar Bose gases
with two- and three-body interactions. 
Furthermore, we have determined corrections to the elementary excitations, quantum and thermal fluctuations of homogeneous dipolar gases
originating from effects of temperature and three-body interactions  using beyond mean field treatment.
We have found that the interplay of the DDI, two- and three-body interactions can sorely modify the sound velocity,
LHY EoS, compressibility and the ground state energy of the system.
The superfluid fraction which becomes an anisotropic quantity by virtue of the DDI is also lowered owing to the nesting effects of thermal fluctuations and three-body interactions.

In addition, we have accurately studied the role of thermal fluctuations on the formation of droplets in Bose gases with strong DDI.
The presence of the GLHY which adds an extra term, inexistent in the standard LHY, to the GP equation neutralizes the dipolar implosion at large density and therefore, 
allowing quantum and thermal fluctuations to form robust droplets at low temperature.
The HFBP model with GLHY term complemented by a direct numerical simulation, has pointed out that at finite temperature,
each droplet forms a thermal halo-like structure. As the temperature approaches the transition, 
this thermal bath inflates and thus, the droplet disappears.

We hope that the insights obtained in this work would help to find a rich set of phases in more strongly dipolar systems, 
such as Rydberg gases \cite{Cinti} or polar molecules.
Our results could also be useful in elucidating the prominent role of the quantum and thermal fluctuations on the formation of a droplet state in 2D dipolar gases \cite{Boudj}.
An interesting future theoretical challenge is to analyze the behavior of the self-bound dipolar droplet in the presence of the anomalous density
using the full HFB theory. The anomalous density could play a key role in understanding the superfluidity of liquid droplets
since both quantities arise from atomic correlations \cite{Boudj3, Boudj4}.      
At large anomalous correlations (pairing instability), one can expect that the self-bound BEC may split 
into several peaks which remain spatially localized \cite{Boudj5, Boudj6}.

\section{Acknowledgments}
We are grateful to Dmitry Petrov, Igor Ferrier-Barbut, Axel Pelster, Lauriane Chomaz and Francesca Ferlaino for the careful reading of the manuscript and helpful comments. 
We thank Hiroki Saito for valuable discussions.

\end{document}